%Paper: hep-ph/9308328
%From: NELLO@TRIESTE.INFN.IT
%Date: Tue, 24 Aug 1993 17:44:41 +0200 (WET-DST)

\magnification=\magstep1
\hsize=6.2truein
\hoffset=0.15truein
\vsize=8.9truein
\voffset=0.056truein
\phantom{.}
\line{\hfil{IC/93/186}}
\line{\hfil{UTS-DFT-93-18}}
\vskip 2cm
\centerline{\bf VECTOR MESON EXCHANGES}
\smallskip
\centerline{\bf AND CP ASYMMETRY IN $K^{\pm}\rightarrow\pi^{\pm}\pi^0$}
\vskip 1.5cm
\centerline{\bf Riazuddin}
\medskip
\centerline{International Centre for Theoretical Physics, Trieste, Italy}
\smallskip
\centerline{and}
\smallskip
\centerline{Department of Physics, King Fahd University of Petroleum and
Minerals,}
\centerline{Dhahran, Saudi Arabia}
\vskip .4in
\centerline{{\bf N. Paver}\footnote{*}{Supported in part by MURST} and
{\bf F. Simeoni}}
\medskip
\centerline{Dipartimento di Fisica Teorica, University of Trieste, Italy}
\smallskip
\centerline{and}
\smallskip
\centerline{Istituto Nazionale di Fisica Nucleare, Sezione di Trieste, Italy}

\vskip 2cm
\centerline{\bf Abstract}
\medskip
\midinsert\narrower\narrower
\noindent Using a current algebra framework, we discuss the contribution
of vector meson
exchanges to the CP violating asymmetry in the decay
$K^{\pm}\rightarrow\pi^{\pm}\pi^0$, resulting from the interference of
the $K\rightarrow\pi\pi$ amplitude with
the radiative correction $K\rightarrow\pi\pi\gamma$.
\endinsert
\vfill\eject
Many efforts have been recently devoted to the study of processes
sensitive to the existence of direct CP violation in electroweak non-leptonic
transition amplitudes [1], with the aim of
formulating alternatives to $K^0\rightarrow\pi\pi$ decay, where the
present experimental situation is still not well settled [2].
Since, unlike the neutral ${\bar K}K$ system, charged kaons cannot mix, charge
asymmetries in $K^{\pm}$ decays should be considered as convenient
observables to provide evidence for direct CP violation. In this regard,
an interesting discussion was presented in Ref.[3], pointing
out the potential role of the CP asymmetry for the decay
${K^{\pm}\rightarrow\pi^{\pm}\pi^0}$:
$${\cal A}_{\pi^{\pm}\pi^0}={\Gamma(\pi^+\pi^0)-\Gamma(\pi^-\pi^0)\over
\Gamma(\pi^+\pi^0)+\Gamma(\pi^-\pi^0)}.\eqno(1)$$
This asymmetry can be induced by the mixing of the $K\rightarrow\pi\pi$ and
$K\rightarrow\pi\pi\gamma$ channels. Also, neglecting other mixings of these
channels, and assuming CPT invariance stating the equality of $K^+$ and
$K^-$ total rates, any prediction for ${\cal A}_{\pi^{\pm}\pi^0}$ can be
translated into a prediction for the CP violating charge asymmetry of the
rare process ${K^{\pm}\rightarrow\pi^{\pm}\pi^0\gamma}$:
$${\cal A}_{\pi^{\pm}\pi^0\gamma}={\Gamma(\pi^+\pi^0\gamma)-
\Gamma(\pi^-\pi^0\gamma)\over
\Gamma(\pi^+\pi^0\gamma)+\Gamma(\pi^-\pi^0\gamma)}={\cal A}_{\pi^{\pm}\pi^0}\
{Br(K^{\pm}\rightarrow\pi^{\pm}\pi^0)\over Br(K^{\pm}\rightarrow
\pi^{\pm}\pi^0\gamma)}.\eqno(2)$$
Using the Standard Model with the CKM
matrix providing the CP violating electroweak phase, and the effective chiral
Lagrangian framework for the matrix elements of ${K\rightarrow\pi\pi}$ and
$K\rightarrow\pi\pi\gamma$, upper limits of the order of $10^{-6}$ and
$10^{-4}$ were obtained
in Ref.[3] for the asymmetries
${\cal A}_{\pi^{\pm}\pi^0}$ and ${\cal A}_{\pi^{\pm}\pi^0\gamma}$
respectively, quite interesting for CP violation searches
both at hadroproduced intense kaon beams and at
$\phi$-factories [4].
Actually, the relevant $\Delta S=1$ operator is $O(p^6)$ in
chiral perturbation theory [5], which is not known in general,
so that only an order of magnitude estimate could be presented in
Ref.[3].\par
Therefore, it should be interesting to try to explore
also some alternative ways of evaluating the hadronic matrix
elements for the transition ${K\rightarrow\pi^{\pm}\pi^0\gamma}$ relevant to
this problem. Specifically, in what follows we shall present a qualitative
estimate of the contribution of vector meson exchanges, using an approach
inspired by current algebra and PCAC.\par
Our discussion will closely follow the presentation of Ref.[3],
and we introduce the diagrams of Figs.$1_{a,b}$, whose interference can give
a non-vanishing asymmetry ${\cal A}_{\pi^{\pm}\pi^0}$.
In these diagrams open and full circles indicate electroweak and
strong-electromagnetic vertices. In the approximation we will adopt in the
following, of neglecting dispersive parts of loop diagrams with virtual photon
and meson exchanges (including the radiative corrections to Fig.$1_a$), the
interference of Fig.$1_a$ with Fig.$1_b$ is easily seen to be the dominant
mechanism for the CP asymmetry in $K^{\pm}\rightarrow\pi^{\pm}\pi^0$ .\par
Denoting by $T_{(0)}$ and $T_{(\gamma)}$ the corresponding amplitudes,
and by $\delta_0^2$ and $\delta_{\gamma}$ the strong interaction rescattering
phases, the asymmetry (1) is expressed as:
$${\cal A}_{\pi^{\pm}\pi^0}={2 Im(T_{(0)}T_{(\gamma)}^*)\over |T_{(0)}|^2}
\sin{(\delta_{\gamma}-\delta_0^2)}.\eqno(3)$$
Eq.(3) shows the need for two interfering transition amplitudes to the same
final state, with different electroweak as well as strong phases.
Taking into account that $T_{(0)}$ is $\Delta I=3/2$, so that
CP violation there is proportional to $\alpha_{e.m.}$, to leading order in
$\alpha_{e.m.}$ the CP asymmetry is determined by the interference of the
CP conserving part of $T_{(0)}$ with the CP violating part of
$T_{(\gamma)}$, which clearly must come from CP violation in
$T(K^{\pm}\rightarrow\pi^{\pm}\pi^0\gamma)$. For the latter we consider the
contribution from the ``contact'' interaction, mediated at the quark level
by the electromagnetic penguin diagram, which may be expected to play a role,
due to the large {\it top} mass allowed by present experimental limits.
Conversely, the bremsstrahlung amplitude
of $K^{\pm}\rightarrow\pi^{\pm}\pi^0\gamma$ is still $\Delta I=3/2$, and
therefore its CP violating part should also be suppressed as a higher power
of $\alpha_{e.m.}$, by the same argument as above.\par
While the value of $T_{(0)}$ can be taken as real, and determined {\it e.g.}
from the experimental $K^{\pm}\rightarrow\pi^{\pm}\pi^0$ branching ratio,
$T_{(\gamma)}$ must be estimated theoretically. For this purpose, as shown
in Fig.$1_b$, one needs to independently
estimate the process $\pi\pi\gamma\rightarrow\pi\pi$ and the electromagnetic
penguin contribution to the decay amplitude $K\rightarrow\pi\pi\gamma$.\par
The low-energy $\pi\pi\gamma\rightarrow\pi\pi$ amplitude should be reasonably
well described by chiral perturbation theory [5] in leading
order, which gives for an on-shell photon with polarization
$\epsilon_{\mu}(q)$:
$$\eqalign{T(\pi^{\pm}\pi^0\gamma\rightarrow\pi^{\pm}\pi^0)=&{2ie\over
3F_{\pi}^2}
\Big[{p\cdot\epsilon\over p\cdot q}\big(p\cdot q-2k\cdot k^{\prime}+
p^{\prime}\cdot(k^{\prime}-k)\big)\cr
&+{p^{\prime}\cdot\epsilon\over p^{\prime}
\cdot q}\big(p^{\prime}\cdot q+2k\cdot k^{\prime}+p\cdot(k-k^{\prime})\big)
\Big],\cr}\eqno(4)$$
with $F_{\pi}=93\ MeV$. Referring to Figs.1$_{a,b}$, we denote by $K$ the
$K^{\pm}$ momentum, by $p^{\prime}$ and $k^{\prime}$ the outgoing
$\pi^{\pm}$ and $\pi^0$ momenta, and by $p$, $k$ and $q$ the intermediate
$\pi^{\pm}$, $\pi^0$ and $\gamma$ momenta, so that ${K=p+k+q=p^{\prime}+
k^{\prime}}$.
\par
As for the electromagnetic penguin diagram contribution to
${K\rightarrow\pi\pi\gamma}$, the relevant quark operator for an on-shell
photon is represented by the magnetic interaction [6]:
$${\cal L}_{EW}={\sqrt 2}G_F{e\over (4\pi)^2}V_{ts}^*V_{td}F_2(x_t)O_M
+h.c.,\eqno(5)$$
where
$$O_M=i(m_s{\bar s}_R\sigma_{\mu\nu}d_L+m_d{\bar s}_L\sigma_{\mu\nu}d_R)
F_{\mu\nu},\eqno(6)$$
with ${\displaystyle L,R={1\over 2}(1\mp\gamma_5)}$, and
${F_{\mu\nu}=-i(q_{\mu}\epsilon_{\nu}-q_{\nu}\epsilon_{\mu})}$
the electromagnetic tensor. Also, in Eq.(5), $F_2(x_t)$ is the Wilson
coefficient (only the {\it top} quark contribution needs be considered here):
$$F_2(x_t)=\left({x_t\over 2(x_t-1)^3}\right)
\left({8x_t^2+5x_t-7\over 6}-{x_t(3x_t-2)\ln{x_t}\over (x_t-1)}\right),
\eqno(7)$$
where ${x_t=m_t^2/M_W^2}$. For simplicity, and because the estimates of
hadronic matrix elements can only be done within approximations,
we neglect QCD corrections to $F_2$. Clearly, the CP
violating electroweak phase needed for a non-vanishing CP asymmetry is
contained in the CKM coefficient in Eq.(5).\par
Neglecting $m_d\ll m_s$, the hadronic matrix element relevant to Eq.(5)
can be written as
$$\langle\pi^{\pm}\pi^0\gamma\vert e\hskip 2pt O_M\vert K^{\pm}\rangle=
e\hskip 2pt m_s\hskip 2pt {\cal E}\left[(k\cdot\epsilon)(K\cdot q)-
(K\cdot\epsilon)(k\cdot q)\right],\eqno(8)$$
where for practical purposes one can neglect the variation of kinematical
variables and consider ${\cal E}$ as a constant.\par
We finally must combine Eqs.(4)-(8), sum over photon polarizations
and evaluate the phase space integral of the resulting function, as required
by Fig.$1_b$. Limiting to the calculation of the
absorptive part, where the three-particle intermediate state is on-shell,
this integration can be done analytically, using kinematical formulae
[7]. Basically, the three-body phase space can be divided into
two two-body phase spaces with the aid of the auxiliary four-vector ${Q=p+k}$
and the variable ${s=Q^2}$:
$$d\Omega(PS)\delta^{(4)}(K-p-k-q)=ds\delta^{(4)}(Q-p-k){d{\vec p}\over
2p_0}{d{\vec k}\over 2k_0}\delta^{(4)}(K-Q-q){d{\vec Q}\over 2Q_0}
{d{\vec q}\over 2q_0}.\eqno(9)$$
The two-body phase space integrations can be easily performed in the
corresponding CM frames, ${{\vec p}+{\vec k}=0}$ and
${{\vec q}+{\vec Q}=0}$ respectively, only one needs
to express the variables which contain both primed and unprimed momenta in
Fig.$1_b$ by means of the Lorents boost connecting the two systems.
The resulting expression for the asymmetry
(1), in the approximation  $m_{\pi}=0$, is the following:
$${\cal A}_{\pi^{\pm}\pi^0}\simeq {7\over 24}\hskip 2pt{m_K^6\over 256\pi^4}
\hskip 2pt\alpha_{e.m.}{G_F\over{\sqrt 2}}
\hskip 2pt{\cal E}\hskip 2pt{m_s F_2(x_t)\over 3F_{\pi}^2T_{(0)}}
\hskip 2ptIm(V_{ts}^*V_{td})\sin{(\delta_{\gamma}-\delta_0^2)}.\eqno(10)$$
A term proportional to $\varepsilon_{\mu\nu\lambda\rho}K_{\mu}k_{\nu}
q_{\lambda}\epsilon_{\rho}$, potentially present in Eq.(8), would lead to
a vanishing integral over phase space, and can thus be omitted. The factor
${7/24}$ in Eq.(10) represents the correction to the approximation, adopted
in Ref.[3], of performing the phase space integral by keeping
the integrand function fixed at its value at $q=0$.\par
Regarding the determination of the coupling constant ${\cal E}$,
in chiral perturbation theory the quark operator $O_M$ of Eq.(6) is of order
$p^6$, and at that order there can be several operators in terms of pion
and kaon fields with the same $SU(3)\times SU(3)$ transformation properties,
and with coupling constants {\it a priori} not fixed by the theory.
In Ref.[3], in the spirit of the chiral Lagrangian framework,
the estimate
$${\cal E}={2\over F_{\pi}^2},\eqno(11)$$
was obtained, leading for maximal rescattering phase difference
$\delta_{\gamma}-\delta_0^2=\pi/2$ to the limits on ${\cal A}_{\pi^{\pm}\pi^0}$
and ${\cal A}_{\pi^{\pm}\pi^0\gamma}$ mentioned above.
\par Turning to the alternative realization of ${\cal E}$ in terms of vector
meson exchanges, the contributions of vector mesons to the left-side of (8)
are represented in Fig.2. Similar to Fig.1, the open circles represent the
electroweak vertex of the operator $O_M$, while the full ones are the strong
vector-meson couplings to pseudoscalar mesons.\par
For the strong couplings in Fig.2 we adopt the conventional notation,
with $i,j,k$ SU(3) indices:
$$\langle M^j(p_1),M^k(p_2)\vert T\vert V^i(P,\eta)\rangle=-if_{ijk}
(p_1-p_2)\cdot\eta\ F_{VPP}.\eqno(12)$$
\noindent Regarding the weak vertices, for convenience we represent
$O_M$ as follows
$$O_M={m_s\over 2}O^{6-i7}\equiv{m_s\over 2}\left[V_{\mu\nu}^{6-i7}+
A_{\mu\nu}^{6-i7}\right]F_{\mu\nu},\eqno(13)$$
with
$$V_{\mu\nu}^i={\bar q}\sigma_{\mu\nu}{\lambda^i\over 2}q;\qquad\qquad
A_{\mu\nu}^i={\bar q}\sigma_{\mu\nu}\gamma_5{\lambda^i\over 2}q,\eqno(14)$$
where $q$ denotes the quark triplet {\it u, d, s} and $\lambda$'s are
Gell-Mann matrices acting on quark flavors.\par
To estimate the vertices $\langle\gamma, M^j\vert e\hskip 2pt
O^{6-i7}\vert V^i\rangle$ in
Fig.2, we first reduce the photon, which gives the factor
${-ie(q_{\mu}\epsilon_{\nu}-q_{\nu}\epsilon_{\mu})}$, and treat the
remaining vector meson-to-pseudoscalar meson matrix elements of
$V_{\mu\nu}^{6-i7}$ and $A_{\mu\nu}^{6-i7}$ using current algebra and PCAC.
By standard soft-meson techniques we thus find
$$\langle\gamma(q,\epsilon),M^j\vert
O^{6-i7}\vert V^i(P,\eta)\rangle\simeq -{i\over F_{\pi}}
\langle 0\vert\left[F_5^j,A_{\mu\nu}^{6-i7}\right]\vert V^i(P,\eta)\rangle
(-ie)(q_{\mu}\epsilon_{\nu}-q_{\nu}\epsilon_{\mu}),\eqno(15)$$
where $F_5^j$ represents the axial charge with the quantum numbers of $M^j$.
The equal-time commutators
of the axial charges with the tensor densities needed in
Eq.(15) can be taken from the quark model:
$$\left[F_5^j,V_{\mu\nu}^l\right]=-d_{jlk}A_{\mu\nu}^k;\qquad\qquad
\left[F_5^j,A_{\mu\nu}^l\right]=-d_{jlk}V_{\mu\nu}^k,\eqno(16)$$
and we define the vacuum-to-vector meson matrix element of the tensor
current density as
$$\langle 0\vert V_{\mu\nu}^k\vert V^i(P,\eta)\rangle=i\delta_{ik}
F_V^T(\eta_{\mu}P_{\nu}-\eta{\nu}P_{\mu}),\eqno(17)$$
with $F_V^T$ a new coupling constant.\par
The value of the tensor constant $F_V^T$ is not obtainable from symmetry
arguments, and thus must be estimated independently. The simplest
possibility is represented by the constituent quark model, which predicts
$$F_V^T={F_V\over M_V},\eqno(18)$$
where $F_V$ is the leptonic vector meson constant, related to $V\rightarrow
e^+e^-$:
$$\langle 0\vert V_{\mu}^k\vert
V^i(P,\eta)=\delta_{ij}F_V\eta_{\mu}.\eqno(19)$$
In Eq.(19),
${\displaystyle V_{\mu}^k={\bar q}\gamma_{\mu}{\lambda^k\over 2}q}$ are
the familiar SU(3) vector currents.\par
To simplify the theoretical expressions, we can apply the vector
dominance determination of the strong coupling constants in Eq.(12):
$$F_Vg_{VPP}=m_V^2.\eqno(20)$$
\par Combining Eqs.(12)-(20), the vector-meson poles of Fig.2 give
the following contribution to the coupling constant ${\cal E}$ of Eq.(8):
$$\eqalign{{\cal E}|_{VM}\cong{2\over F_Km_{\rho}}
\Bigg[&{m_{\rho}^2\over m_{\rho}^2+(K-q)^2}\cr
&+{1\over 2}{F_K\over F_{\pi}}{m_{\rho}\over m_{K^*}}
\left({m_{K^*}^2\over m_{K^*}^2+(K-k)^2}
+{m_{K^*}^2\over m_{K^*}^2+(K-p)^2}\right)\Bigg],\cr}\eqno(21)$$
to be compared with Eq.(11). In Eq.(21), we have phenomenologically included
some SU(3) symmetry breaking effects, although the derivation is such that,
in principle, all quantities there should be evaluated in the symmetry
limit. As stated above, for our purpose, to a good approximation we can
neglect the polar dependence of the denominators in Eq.(21).\par
It might be interesting to notice that the same result as in Eq.(21)
can also be directly obtained
by expanding the matrix element ${\langle\pi^{\pm}\pi^0\gamma\vert
O_M\vert K^{\pm}\rangle}$ needed in (8), to the leading order in the photon and
meson four-momenta, by means of the reduction formalism and soft pion and
kaon techniques. Firstly, by reducing the photon and the external
pions and using PCAC Ward identities and current algebra commutators
[8], one can derive the expansion (double
equal-time commutators, which appear in general, are easily seen to have
vanishing matrix elements in this specific case\footnote{$\dagger$}{The same
is true for the (esplicitly O($m_{\pi}^2$)) $\sigma$-term}):
$$\eqalign{F_{\pi}^2&\langle\pi^+\pi^0\gamma
\vert e\hskip 2pt O_M\vert K^+(K)\rangle
\cong (-ie)\left(q_{\lambda}\epsilon_{\nu}-q_{\nu}\epsilon_{\lambda}\right)
m_s \Big[-{i\over 2{\sqrt 2}}\left(p-k\right)_{\mu} M_{\mu\lambda\nu}\cr
&+i{F_{\pi}\over{\sqrt 2}}\langle\pi^0(k)\vert V_{\lambda\nu}^{4-i5}(0)
\vert K^+(K)\rangle -i{F_{\pi}\over 2}
\langle\pi^+(p)\vert V_{\lambda\nu}^{6-i7}(0)\vert K^+(K)\rangle\Big],
\cr}\eqno(22)$$
where
$$M_{\mu\lambda\nu}=\int d^4x\exp{[-i(k+p)\cdot x]}
\langle 0\vert T\left(V_{\mu}^{1-i2}(x)A_{\lambda\nu}^{6-i7}(0)
\right)\vert K^+\rangle.\eqno(23)$$
Furthermore, one can still expand Eq.(23) by reducing the
external kaon, and using the quark model equal-time commutators (16) in
local form (derivative terms, such as Schwinger terms should contribute higher
powers of the momenta). One finds
$$M_{\mu\lambda\nu}\simeq {i\over{\sqrt 2} F_K}
\left[\Delta_{\mu\lambda\nu}^{\rho^+}(k+p)+
\Delta_{\mu\lambda\nu}^{K_A^0}(-q)\right]
-iK_{\rho}M_{\mu\lambda\nu\rho},\eqno(24)$$
with
$$\eqalign{M_{\mu\lambda\nu\rho}={1\over\sqrt 2}  \int\int d^4x d^4y &
\exp{[-i(k+p)\cdot x]}\exp{i(K\cdot y)}\cr
&\times\langle 0\vert T
\left(V_{\mu}^{1-i2}(x)A_{\lambda\nu}^{6-i7}(0)A_{\rho}^{4+i5}(y)\right)
\vert 0\rangle,\cr}\eqno(25)$$
where
${\displaystyle A_{\rho}^k={\bar q}\gamma_{\rho}\gamma_5{\lambda^k\over 2}q}$.
In Eq.(24) $q=K-p-k$, and we have introduced the two-point spectral functions:
$$\Delta_{\mu\lambda\nu}^{\rho^+}(P)=\int d^4x
\exp{(-iP\cdot x)}\ \langle 0\vert T\left(V_{\mu}^{1-i2}(x)
V_{\lambda\nu}^{1+i2}(0)\right)\vert 0\rangle,\eqno(26)$$
$$\Delta_{\mu\lambda\nu}^{K_A^0}(P)=\int d^4x
\exp{(-iP\cdot x)}\ \langle 0\vert T\left(A_{\mu}^{6+i7}(x)
A_{\lambda\nu}^{6-i7}(0)\right)\vert 0\rangle.\eqno(27)$$
Due to the vector current conservation, the surface term
$K_{\rho}M_{\mu\lambda\nu\rho}$ is of higher order in momenta.
\par Similarly, one can expand the last two matrix elements on the
right-side of (22) to leading order in the pion and kaon momenta (equal-time
commutators matrix elements vanish also here):
$$\eqalign{\langle\pi^0(k)&\vert V_{\lambda\nu}^{4-i5}\vert K^+(K)\rangle\cr
&\simeq{-i(k+K)_{\mu}\over 4{\sqrt 2}F_KF_{\pi}}\int d^4x
\exp{[-i(k-K)\cdot x]}\ \langle 0\vert T\left(V_{\mu}^{4+i5}(x)
V_{\lambda\nu}^{4-i5}(0)\right)\vert 0\rangle,\cr}\eqno(28)$$
and
$$\eqalign{\langle\pi^+(p)&\vert V_{\lambda\nu}^{6-i7}\vert K^+(K)\rangle\cr
&\simeq{-i(p+K)_{\mu}\over 4F_KF_{\pi}}\int d^4x
\exp{[-i(p-K)\cdot x]}\ \langle 0\vert T\left(V_{\mu}^{6+i7}(x)
V_{\lambda\nu}^{6-i7}(0)\right)\vert 0\rangle.\cr}\eqno(29)$$
Replacing these expansions into Eq.(22), the axial-vector spectral function
$\Delta_{\mu\lambda\nu}^{K_A^0}(-q)$ does not contribute after contraction of
Lorentz indices, so that finally the matrix element of interest reduces to
the following combination:
$$\eqalign{F_{\pi}^2\hskip 2pt{\cal E}\cong {1\over 4F_K}
\Big[&(p-k)_{\mu}\Delta_{\mu\lambda\nu}^{\rho^+}(p+k)\cr
&+{1\over 2}(k+K)_{\mu}\Delta_{\mu\lambda\nu}^{K^{*-}}(k-K)-
{1\over 2}(p+K)_{\mu}
\Delta_{\mu\lambda\nu}^{K^{*0}}(p-K)\Big].\cr}\eqno(30)$$
In Eq.(30), $\Delta_{\mu\lambda\nu}^{K^{*-}}$ and
$\Delta_{\mu\lambda\nu}^{K^{*0}}$ are defined analogously to
$\Delta_{\mu\lambda\nu}^{\rho^+}$, by obvious replacement of the $SU(3)$
quantum numbers according to (28) and (29). Vector-meson saturating the
spectral functions in (30), and using Eqs.(18)-(20) and the relation [9]
${\displaystyle F_{K^*}^2=2F_{\pi}F_Km_{K^*}^2}$, directly leads to
the same result for the coupling constant ${\cal E}$ as in Eq.(21).\par
Using an upper limit on $Im(V_{ts}^*V_{td})/V_{us}V_{ud}$ of
$1\times 10^{-3}$ from recent fits to the CKM matrix [10],
with $|V_{ud}|=0.975$ and $|V_{us}|=0.221$, and the values of quark masses
$m_s=200\ MeV$ and $m_t=150\ GeV$, for maximal rescattering relative phase
we would find from Eqs.(10) and (30):
$${\cal A}_{\pi^{\pm}\pi^0}\vert_{VM}<2\times 10^{-8}\hskip 2pt;\qquad\qquad
{\cal A}_{\pi^{\pm}\pi^0\gamma}\vert_{VM}< 1.5\times 10^{-5}.\eqno(31)$$
\par The values in Eq.(31) are somewhat smaller than those found in
Ref.[3], reported at the beginning. This reduction is due
to the joint effects of the suppression factor $7/24$ of the phase space
integration in Eq.(10), of the scale of the constant ${\cal E}$ set in
Eq.(21) by the
vector meson mass, which suppresses its value relative to Eq.(11) by a
factor $2F_{\pi}/m_{\rho}$, and finally to the more severe constraints on
{\it top} CKM angles used here.\par
The size of CP violation in Eq.(31) could compete with that computed in
[11], by considering the contribution of the gluon-penguin operator to the
electric-dipole transition.
\vskip 5cm
\centerline{\bf Acknowledgements}
\smallskip
One of the authors (R) would like to acknowledge the support of the King Fahd
University of Petroleum and Minerals, and of INFN-Sezione di Trieste. He
also thanks Prof. Abdus Salam, the International Atomic Energy Agency and
Unesco for hospitality at the International Centre for Theoretical Physics,
Trieste. The other author (NP) is grateful to Dr. A.Ali and Dr. G.D'Ambrosio
for useful discussions.
\vfill
\eject
\vskip 1in
\centerline{\bf References}
\vskip 0.5in
\item{[1]} For references see e.g: {\it CP-Violation}, Ed. L.Wolfenstein,
North-Holland, Amsterdam, 1989; {\it CP-Violation}, Ed. C.Jarskog,
World Scientific, Singapore, 1989.
\medskip
\item{[2]} G.Barr, NA31 Collaboration, in {\it Proceedings of the Joint
International Lepton-Photon Symposium \& Europhysics Conference on High
Energy Physics}, Geneva, Switzerland, 1991, Eds. S.Hegarty, K.Potter and
E.Quercigh, World Scientific (1992), p. 179; B. Winstein, E731 Collaboration,
{\it ibidem}, p. 186.
\medskip
\item{[3]} C.O.Dib and R.D.Peccei, Phys. Lett. {\bf B249} (1990) 325.
\medskip
\item{[4]} See e.g. {\it The DAFNE Physics Handbook}, Eds. L.Maiani,
G.Pancheri and N.Paver (1992).
\medskip
\item{[5]} See for instance H. Georgi, {\it Weak Interactions and Modern
Particle Theory}, Benjamin/Commings, Menlo Park 1984; J.Gasser and
H.Leutwyler, Ann. Phys. {\bf 158} (1984) 142, Nucl. Phys. {\bf B250} (1985)
465.
\medskip
\item{[6]} T.Inami and C.S.Lim, Prog. Theor. Phys. {\bf 65} (1981) 297,
1772 (E).
\medskip
\item{[7]} B.R.Martin, E. de Rafael and J.Smith, Phys. Rev. D {\bf 2} (1970)
179.
\medskip
\item{[8]} See, e.g.: R. Marshak, Riazuddin and C. P. Ryan, ``{\it Theory
of Weak Interactions in Particle Physics}'', John Wiley (1969),
p. 491.
\medskip
\item{[9]} K.Kawarabayashi and M.Suzuki, Phys. Rev. Lett. {\bf 16} (1966) 255;
Riazuddin and Fayyazuddin, Phys. Rev.{\bf 147} (1966) 1071.
\medskip
\item{[10]} A.Ali and D.London, report DESY 93-022 (1993).
\medskip
\item{[11]} M.McGuigan and A.I.Sanda, Phys. Rev. D {\bf 36} (1987) 1413.
\vfill\eject
\vskip 1cm
\centerline{\bf Figure captions}
\vskip 0.5in
\noindent Fig.1$_{a,b}$: Amplitudes contributing to the asymmetry Eq.(1).
\medskip
\noindent Fig.2: Vector meson exchanges contributing to Eq.(8).
\vfill
\eject
\bye